\documentclass[a4paper,11pt]{article}
\pdfoutput=1 
\usepackage{appendix}
\usepackage{jcappub} 

\usepackage[T1]{fontenc} 
\usepackage[style=numeric,backend=biber, sorting = none]{biblatex}

\title{\boldmath A title with some math: $x=1$}

\newcommand{\mpcoh}{\,h^{-1}\,{\rm Mpc}}

\newcommand{\oiii}{[O{\sc\,iii}]}
\newcommand{\oii}{[O{\sc\,ii}]}

\title{\textbf{Correcting for small-displacement interlopers in BAO analyses}}

\author[a,b]{Setareh Foroozan}
\author[a,b]{Elena Massara}
\author[a,b,c]{Will J. Percival}

\affiliation[a]{Waterloo Centre for Astrophysics, University of Waterloo, 200 University Ave W, Waterloo, ON N2L 3G1, Canada}
\affiliation[b]{Department of Physics and Astronomy, University of Waterloo, 200 University Ave W, Waterloo, ON N2L 3G1, Canada}
\affiliation[c]{Perimeter Institute for Theoretical Physics,
31 Caroline St. North, Waterloo, ON N2L 2Y5, Canada}

\emailAdd{s2forooz@uwaterloo.ca}

\abstract{
Due to the low resolution of slitless spectroscopy, future surveys including those made possible by the Roman and Euclid space telescopes will be prone to line mis-identification, leading to interloper galaxies at the wrong redshifts in the large-scale structure catalogues. The most pernicious of these have a small displacement between true and false redshift such that the interloper positions are correlated with the target galaxies. We consider how to correct for such contaminants, focusing on $\rm H\beta$ interlopers in [O{\sc\,iii}] catalogues as will be observed by Roman, which are misplaced by $\Delta d = 97 \,h^{-1}\,{\rm Mpc}$ at redshift $z = 1$. Because this displacement is close to the BAO scale, the peak in the interloper-target galaxy cross-correlation function at the displacement scale can change the shape of the BAO peak in the auto-correlation of the contaminated catalogue, and lead to incorrect cosmological measurements if not accounted for properly. We consider how to build a model for the monopole and quadrupole moments of the contaminated correlation function, including an additional free parameter for the fraction of interlopers. The key input to this model is the cross-correlation between the population of galaxies forming the interlopers and the main target sample. It will be important to either estimate this using calibration data or to use the contaminated small-scale auto-correlation function to model it, which may be possible if a number of requirements about the galaxy populations are met. We find that this method is successful in measuring the BAO dilation parameters without significant degradation in accuracy, provided the cross-correlation function is accurately known.}

\begin{document}
\maketitle
\flushbottom

\section{Introduction}\label{sec:intro}

Space-based Large Scale Structure (LSS) spectroscopic surveys made possible by the Roman Space Telescope \cite{spergel2013widefield} and Euclid \cite{laureijs2011euclid} missions will soon be available. These surveys will make use of slitless spectroscopy to measure galaxy redshifts meaning that, while they have the potential to constrain the cosmological parameters with sub percent levels of accuracy, this requires a thorough understanding of potential sources of systematic error arising from the slitless method. One possible systematic effect, that we study in this paper, is the contamination of the sample by interlopers resulting from line confusion.

To estimate the redshifts of a set of galaxies using slitless spectroscopy, we measure the wavelengths of one or more emission lines for each. In this paper we will primarily be concerned with samples where only one emission line is used to estimate the redshift. Generally, emission lines with different rest-frame wavelengths $\lambda_1$ and $\lambda_2$, from two different galaxies with true redshifts $z_1 = \lambda_{\rm obs} / \lambda_{1}-1$ and $z_2 = \lambda_{\rm obs} / \lambda_{\rm 2}-1$, can appear at the same observed wavelength, $\lambda_{\rm obs.}$. The measured redshift for a galaxy given only the observed position of this single line will depend on the rest-frame wavelength assumed. In the presence of a secondary emission line, one can distinguish between different options. However, if we are measuring redshifts from a single emission line, we are prone to line misidentification, if we assume, for a sample of galaxies, that the observed line has the same rest-frame wavelength for all galaxies. The galaxies in the sample for which the lines are misidentified resulting in them being assigned the wrong redshift, are called interlopers. We refer to the galaxies whose observed emission line matches the assumed rest-frame wavelength when measuring redshifts as the target galaxies. 

For Roman, for example, it is likely that many redshifts will be measured using 2 or more emission lines \cite{wang_high_2022}. At $2<z<3$, \oiii\ is the primary line for Roman observations, which is actually a doublet. Resolving this doublet as two separate lines, which is possible at given the resolution of Roman, or observing a secondary line from \oii\ emission would allow a secure redshift to be measured. However, to provide a bonus sample to higher number densities, it will be useful to analyse galaxies where only one unresolved line is observed, assumed to be \oiii, but where this identification is less secure. For this, we will have interlopers, where the H$\beta$ line is mistaken for \oiii. The increased size of this bonus sample will help reduce the statistical errors on measured cosmological parameters, at the cost of increasing the systematic errors. It is these systematic errors, and methods to mitigate them, that we consider in this paper.

In general, there are three different ways that a catalogue can be contaminated: 
\begin{enumerate}
    \item First, due to noise spikes being misidentified as emission lines. The noise is uncorrelated with all the other galaxies, and has a distribution different from that of the true observed lines. Thus, this population of contaminants Poisson samples a volume that has a window different from that of the target galaxy population. It can trivially be corrected in a clustering measurement as it is absorbed into (and is cancelled by) part of the expected density (often quantified by a random catalog) and simply adds to the shot noise and changes the amplitude of the signal measured.
    \item Second, misidentified emission lines can arise from interloper galaxies displaced by a large radial distance. In this case, correlations with the target galaxy population are small and we only need to model the auto-correlation of the interloper sample (e.g. \cite{Farrow_2021,Grasshorn_Gebhardt_2019,Gong_2021,pullen_interloper_2016}).
    Because the displacement is large, there is a significant change in $H(z)$ between the true redshift of the interlopers and the incorrect redshift inferred, such that the Baryon Acoustic Oscillation (BAO) signal in the auto-correlation of the interlopers is shifted with respect to its true value. This needs to be included in any analysis but, as the shift is known from the line emission wavelengths, this contribution to the clustering of the contaminated sample can be easily modeled provided the fraction of interlopers is known.
    \item The third class of contaminant is a more pernicious type of interloper, which is one only shifted by a small radial displacement by the incorrect rest-frame wavelength assignment. Such interlopers will be correlated with other such objects {\em and} with the target galaxy sample. This type of interloper can occur for samples obtained using the Roman Space Telescope at high redshifts, where the target is the primary \oiii\ line, and $\rm H \beta$ line is the source of interlopers. The impact of these interlopers on the BAO peak position was studied by \cite{massara2020line}. They showed that misidentifying $\rm H \beta$ line as \oiii\ line in an Emission Line galaxy will lead to underestimating its distance by $\sim 90\mpcoh$. This would distort the shape of the galaxy correlation function and therefore introduce a shift in the BAO peak position. We call this class of interlopers, which are displaced by $\lesssim 150 \mpcoh$, "small-displacement" interlopers (see Section~\ref{sec:results}).
\end{enumerate}

We focus on the third class of contaminant. Following \cite{massara2020line}, we present a model for the monopole and the quadrupole moments of the correlation function of a Roman-like catalogue, with an \oiii\ target line sample contaminated by $\rm H\beta$ interlopers. With this model, we investigate how to find an unbiased estimation of the isotropic and anisotropic dilation parameters $\alpha$ and $\epsilon$ from the BAO signal, as well as the fraction of interlopers $f_{\rm i}$. Our method is based on standard BAO modeling, additionally allowing for the interlopers. The model correlation function has two terms: The galaxy auto-correlation that we estimate using CAMB \cite{lewis_efficient_2000}, and a non-negligible cross-correlation between the galaxies and the interlopers that we consider a number of ways of estimating. If we can measure this from the contaminated auto-correlation, or use another way of accurately calibrating this statistic, our pipeline gives an unbiased estimation of all the three  parameters. If we do not take into account the interlopers, the measurements would be highly biased.

The plan of our paper is as follows: 
In Section~\ref{subsec:displacement}, we derive the interloper shift value.
In Section~\ref{sec:simulations}, we describe simulations that were used for our analysis.
In Section~\ref{sec:Modelling_the_contaminated_correlation_function}, we present our model for the correlation function of a catalogue contaminated by some small-displacement interlopers. We show the outcomes of our pipeline in Section~\ref{sec:results}. We discuss our results and their implications in Section~\ref{sec:discussions}.

\begin{figure}[t]
    \centering
    \includegraphics[width=1\textwidth,angle=0]{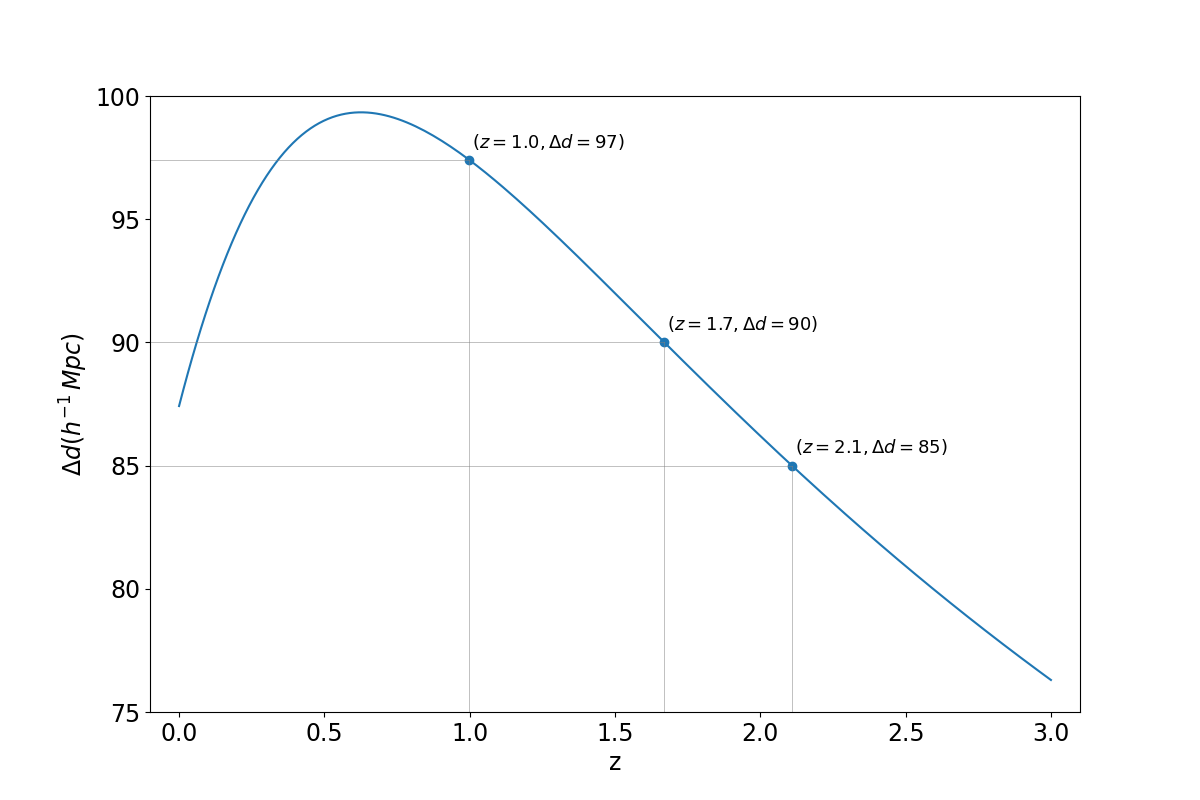}
    \caption{
    Displacement of interlopers as a function of their true redshift as shown in Eq. \ref{eq:displacement}.
    }
    \label{fig:displacement}
\end{figure}

\section{Displacement}\label{subsec:displacement}

Let us assume that in a survey, potential interlopers are at redshift $z^{\rm true}$, and they emit a primary emission line with rest-frame wavelength $\lambda_{\rm emit}^{\rm true}$, that is observed at wavelength $\lambda_{\rm obs}$. The observed wavelength can be misidentified for another photon as if it was emitted with rest-frame wavelength $\lambda_{\rm emit}^{\rm false}$ at redshift $z^{\rm false}$. This redshift misidentification would effectively lead us to a wrong estimation of the comoving proper distance to the galaxies by $\Delta d$:
\begin{subequations}
\label{eq:y}
\begin{align}
    \label{eq:y:1}
        \Delta d =
        d^{\rm true} - d^{\rm false}
        = \int_{z^{\rm false}}^{z^{\rm true}} \frac{cdz}{H(z)}
        \\
    \label{eq:y:2}
        \approx 
        \frac{c}{H(z^{\rm true})} (z^{\rm true} -z^{\rm false})\,.
        \\
    \label{eq:y:3}   
        \approx 
        \frac{c}{H(z^{\rm true})} \left[1 -  \frac{\lambda^{\rm true}_{\rm emit}}{\lambda^{\rm false}_{\rm emit}} \right] (1 + z^{\rm true})\,.
\end{align}
\end{subequations}
In this paper we consider $\rm H \beta$ interlopers misidentified as \oiii\ emitters. While \oiii\ is a doublet, here we assume that $\rm H \beta$ is misidentified as the primary line in the doublet at $\lambda^{\rm OIII}_{\rm false} = 500.7 \, \rm nm$, rather than the secondary line at $\lambda^{\rm OIII}_{\rm false} = 495.9 \, \rm nm$. The rest-frame wavelength of $\rm H \beta$ is $\lambda^{\rm  H\beta}_{\rm true} = 486.1 \,\rm nm$, leading to an offset between the true and wrongly inferred interloper comoving position equal to  
\begin{equation}
\label{eq:displacement}
    \Delta d \approx 87.41 \frac{1+z^{\rm true}}{\sqrt{\Omega_{\Lambda} + \Omega_{\rm m} (1+z^{\rm true})^3}}\, \mpcoh.
\end{equation}
Figure~\ref{fig:displacement} displays the distance $\Delta d$ as a function of the true galaxy redshift and with the parameters $\Omega_\Lambda$ and $\Omega_{\rm m}$ evaluated at the cosmology of the simulations used in this paper and described in Section~\ref{sec:simulations}. In that cosmology, the displacement $\Delta d$ is equal to $97 \mpcoh$ at $z = 1$ (the redshift of the N-body snapshots considered in this work), and $90, 85\mpcoh$ at $z = 1.7, 2.1$, which are the redshifts of interest for the Roman Space Telescope. 

Note that measurements stemming from rest-frame wavelength differences cannot be used as standard rulers in the same way as BAO - what is fixed is the difference in redshift, not the distance. The displacement in physical units depends on the assumed fiducial cosmology used to convert redshifts to distances not the true cosmology, and we do not have to vary the displacement when fitting different cosmological models to the data.

\begin{figure}[ht]
    \centering
    \includegraphics[width=1\textwidth,angle=0]{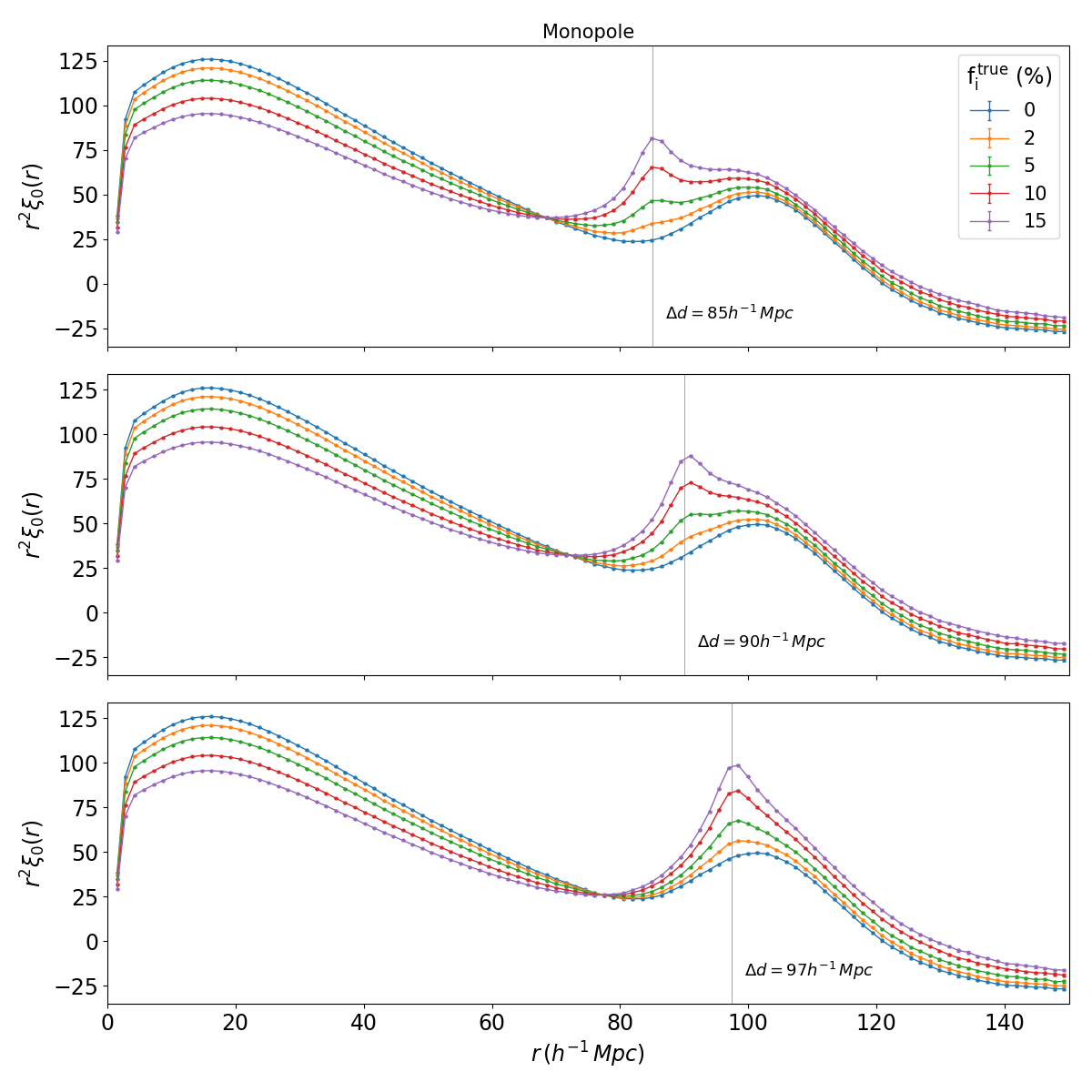}
    \caption{
    Monopole of the correlation function as calculated from the mean of 1000 mocks that are contaminated by different fractions of interlopers (different colours), and different interloper displacements: $85$ (top row), $90$ (middle row), and $97\mpcoh$ (bottom row).
    As can be seen, the BAO peak is skewed towards $\Delta d$, and amplified more as the number of interlopers increase.
    }
    \label{fig:mono_general}
\end{figure}

\begin{figure}[t]
    \centering
    \includegraphics[width=1\textwidth,angle=0]{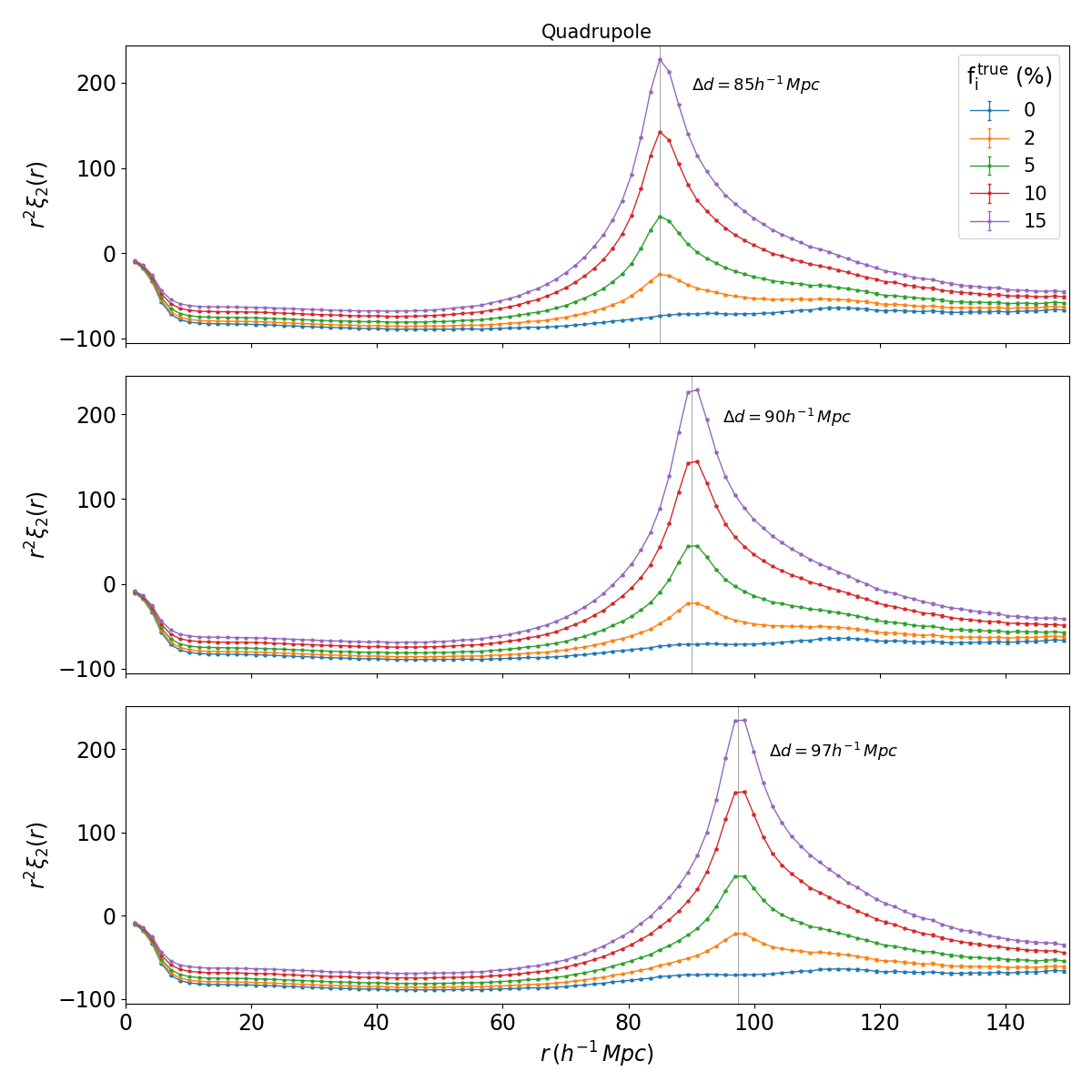}
    \caption{
    Quadrupole of the correlation function as calculated from the mean of 1000 mocks that are contaminated by different fractions of interlopers (different colours), and different interloper displacements: $85$ (top row), $90$ (middle row), and $97\mpcoh$ (bottom row).
    As can be seen, the quadrupole peaks at $\Delta d$, which demonstrates the anisotropy caused by the shift of interlopers along the LOS.
    }
    \label{fig:quad_general}
\end{figure}

\section{Simulations}
\label{sec:simulations}
In this study, our focus is on building and testing a model framework that describes the effect of interloper galaxies rather than predicting the actual expected fraction of interlopers.
While some progress have been made in modelling the population of \oiii \ emitters (e.g., \cite{Zhai_2019}), 
we need a thorough comparison with observations to know if we have the accuracy required. The population of $\rm H \beta$ emitters is less well understood.
Therefore, we will use halos in place of galaxy catalogs to test our model, for simplicity. We employ the halo catalogs from $1,000$ N-body simulations of the Quijote suite \cite{Quijote_2020}. These simulations have been run in a box with size $1 h^{-1}$Gpc and follow the evolution of $512^3$ cold dark matter particles in the fiducial cosmology of the suite, which is a flat $\rm \Lambda CDM$ cosmology with parameters $\Omega_m = 0.3175$, $\Omega_b = 0.049$, $n_s = 0.9624$, $h = 0.6711$, $\sigma_8 = 0.834$, $M_\nu = 0.0$ eV. Halos have been identified using the Friends-of-Friends algorithm with linking length parameter equal to 0.2, and only those with 20 or more cold dark matter particles have been stored in the catalogue. This means that considered halos have minimum mass equal to $1.31 \cdot 10^{13}h^{-1}$M$_{\odot}$. 

Even though \oiii\ emission galaxies are the primary target at redshifts beyond $z=1.8$ in the High-Latitude Survey of the Roman telescope, we use halo catalogs at redshift $z=1$ to have a larger number density of objects, $\bar{n}_h = 2\cdot 10^{-4}h^3{\rm Mpc}^{-3}$, consistent with the expected \oiii\ mean number density in Roman between redshifts $1.8-2.8$  \cite{Zhai_2019,massara2020line}. At the redshift of the simulation, interlopers appear to be displaced by $97 \mpcoh$ along the line of sight. To simulate interlopers at higher redshift, we also consider displacements equal $90$ or $85 \mpcoh$ that correspond to catalogs at redshifts $z=1.7$ and $z=2.1$, respectively. We build the contaminated halo catalogs by randomly selecting objects that will be interlopers and displacing them at the wrongly inferred distance along the $z$ direction, that we assume to be the line-of-sight (hereafter, LOS). We implement four different percentages of interlopers (2, 5, 10, 15 \%) and three different values for the displacement, obtaining 12 different contaminated halo catalogs for each of the initial uncontaminated ones.  

We measure the correlation function from all these mocks, by making use of Nbodykit \cite{Nbodykit_Hand_2018} adopting the Landy-Szalay estimator \cite{landy_szalay_1993}. 
To build a better understanding of the problem, the monopole and quadrupole of this measured correlation function are shown in Figure~\ref{fig:mono_general} and Figure~\ref{fig:quad_general}, respectively. 
In these plots, we show catalogues contaminated by different fractions of interlopers in different colours, and with different displacements in different panels: $\Delta d = 85$ (top), $90$ (middle), and $97 \mpcoh$ (bottom).
The effects of the interlopers are clear to see in both moments. The amplitude of the small-scale monopole is suppressed by the presence of interlopers, and the suppression increases with the fraction of interlopers. The missing small-scales correlation is converted into larger scales, and in particular it enhances the amplitude of the correlation around the displacement scale, which is close to the BAO peak position. As a result, the BAO peak appears to be enhanced, broadened and shifted towards smaller separations. The quadupole is also modified by the interlopers: Its amplitude is enhanced on all scales with increasing interloper fraction. Moreover, it also exhibits a peak around the scales corresponding to the displacement of the interlopers, as a consequence of the increased number of interloper-target galaxy pairs along the LOS.

\section{Modelling the Contaminated Correlation Function}
\label{sec:Modelling_the_contaminated_correlation_function}

In this Section, we show how to model the monopole and quadrupole moments of the correlation function for a catalogue that is contaminated by interlopers misplaced by relatively small displacements such that the interloper-galaxy cross-correlation term is not negligible. We build a model for the auto-correlation function of a contaminated catalogue in Section~\ref{subsec:correlation_function}, and in Section~\ref{subsec:cross_correlation} we introduce our method to estimate the interloper-target cross-correlation function. Finally, in Section~\ref{subsec:Model_and_Parameters}, we discuss our completed model.

\subsection{Correlation Function with Interlopers}\label{subsec:correlation_function}

We consider a catalogue contaminated by a given fraction of interlopers, $f_{\rm i}$, and define $\xi_{\rm gg}$ the auto-correlation function of the target galaxies, $\xi_{\rm gi}$ the cross-correlation between these galaxies and interlopers and $\xi_{\rm ii}$ the auto-correlation function of the interlopers. Throughout this paper, indices "$\rm i$", and "$\rm g$", stand for interlopers, and galaxies respectively.

We denote the observed comoving distance of an object with $\Vec{x}$: for target galaxies, this is the same as their true position, whereas for interlopers it is the wrong measured position. We denote the true position of an interloper with $\Vec{y}$. Following \cite{Pullen2015}, the contaminated galaxy overdensity is given by
\begin{equation}
\label{eq:total_overdensity}
\delta (\Vec{x}) =
    (1 - f_{\rm i}) \delta_{\rm g} (\Vec{x}) + f_{\rm i} \delta_{\rm i}(\Vec{x})\,.
\end{equation}
We can use this equation to calculate the contaminated correlation function at any observed separation $\Vec{r} = \Vec{x}_1 - \Vec{x}_2$,
\begin{equation}
\label{eq:xi_tt_general}
\begin{split}
    \xi(\Vec{x}_1-\Vec{x}_2; f_{\rm i}) &= 
    \langle 
    \delta(\Vec{x}_1) \delta^{*}(\Vec{x}_2) 
    \rangle
    \\
    &=\langle 
    \left[(1 - f_{\rm i}) \delta_{\rm g} (\Vec{x}_1) + f_{\rm i} \delta_{\rm i}(\Vec{x}_1)
    \right]
     \left[
    (1 - f_{\rm i}) \delta^{*}_{\rm g} (\Vec{x}_2) + f_{\rm i} \delta^{*}_{\rm i}(\Vec{x}_2)
    \right]
    \rangle\\
    &= (1-f_{\rm i})^2\xi_{\rm gg}(\Vec{r})+f_{\rm i}^2\xi_{\rm ii}(\Vec{r})+2f_{\rm i}(1-f_{\rm i})\xi_{\rm gi}(\Vec{r})\,.
\end{split}
\end{equation}
This equation can be significantly simplified if we can make two key assumptions. 

First, we assume that the displacement $\Delta d$ in Eq.~\ref{eq:displacement} between the true position of interlopers and the wrongly inferred one is a constant within a redshift bin when calculating the correlation function at a fixed cosmology. Indeed, we only consider correlation functions at separations $r$ smaller than $150\mpcoh$ in the following analysis and in this range the displacement only varies by at most 0.8\% (see Figure~\ref{fig:displacement}). Therefore, we can assume that each object in the pair is displaced by the same amount. Moreover, it is commonly assumed that within a redshift bin $H(z)$ is a constant, therefore, we can assume that the displacement is fixed across the redshift bin considered for all the pairs.
Thus, the wrongly inferred position of each interloper can be written in terms of a constant displacement along the LOS $\hat{z}$: $\vec{x}_{\rm i} = \vec{y}_{\rm i} - \Delta d \, \hat{z}$, where $\vec{y}_{\rm i}$ is the true position.

The interloper auto-correlation function can then be written as $\xi_{\rm ii}(\Vec{x}_1 - \Vec{x}_2) = \xi_{\rm ii}(\Vec{y}_1 - \Vec{y}_2)$ (where we have assumed that the LOS is the same for both galaxies in a pair), because two point correlators are invariant under translations -- they depend only on the relative distance between two objects. This means that the correlation function of the interlopers at the wrong redshifts is equal to the correlation function of the interlopers at their true position.

Interlopers and targets belong to two different populations of galaxies and have in principle different bias schemes. Consequently, we either need calibration data to measure $\xi_{\rm gi}(\Vec{r})$ and $\xi_{\rm ii}(\Vec{r})$ (we discuss this particular case in Section~\ref{sec:discussions}), or we need a way to model this function. On the other hand, we note that since displacements are small, interlopers and targets both trace a matter field with similar amplitude of clustering. It is also likely that interlopers have similar properties and intrinsic clustering as the target galaxies, meaning that they share the same galaxy bias. 
We thus make a second assumption: targets and interlopers share the same galaxy bias. In this case, together with the assumption of constant displacement for the interlopers, we have: 
$\xi_{\rm ii}(\Vec{r}) = \xi_{\rm gg}(\Vec{r})$.
We assume that this is true in the rest of the paper. Our results and conclusions will be similar if this is not true, but instead we are able to accurately estimate or model the bias scheme of the interlopers. Substituting into Eq.~\ref{eq:xi_tt_general} yields
\begin{equation}
\label{eq:xi_tt}
    \xi(\Vec{r};f_{\rm i}) = (1+2f_{\rm i}^2-2f_{\rm i})\xi_{\rm gg}(\Vec{r})+2f_{\rm i}(1-f_{\rm i})\xi_{\rm gi}(\Vec{r})\,.
\end{equation}
In the next Section, we will further consider methods to estimate the cross-correlation between interlopers and galaxies, $\xi_{\rm gi}(r)$.

\begin{figure}[t]
    \centering
    \includegraphics[width=1\textwidth,angle=0]{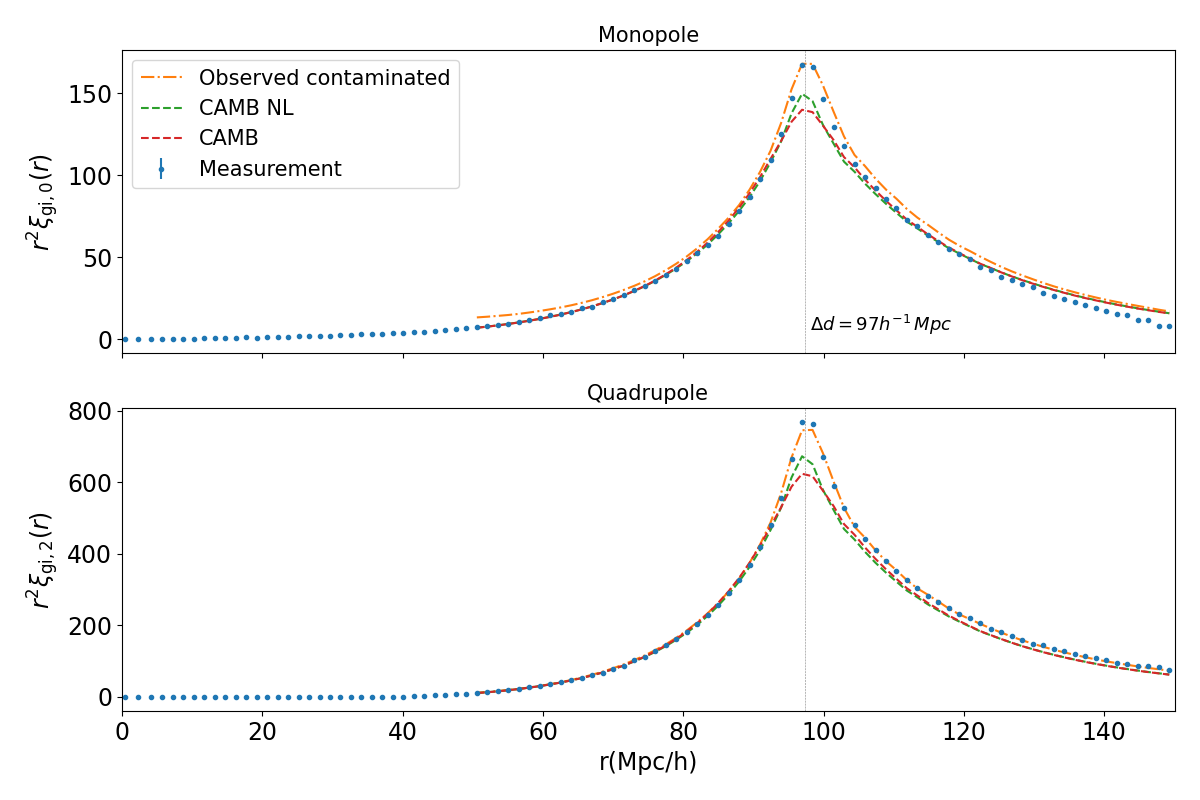}
    \caption{The importance of accurately modelling the small-scale cross-correlation in order to match the large-scale clustering around the BAO feature.
    Monopole (top) and quadrupole (bottom) of the galaxy-interloper cross-correlation function correlation in real-space, calculated from mapping of auto-correlation described in Eq.~\ref{eq:cross_equation}. We considered four different auto-correlations described in more detail in text. The interloper displacement, $\Delta d$, used in mapping equals $97$ for $z = 1$.
    }
    \label{fig:cross_correlation}
\end{figure}

\subsection{Interloper-Target Cross-correlation}
\label{subsec:cross_correlation}
As mentioned before, the cross-correlation between galaxies and interlopers plays a key role in determining the shape of the contaminated correlation function. 
While the galaxy-galaxy and interloper-interloper terms remain unchanged after misplacing interlopers along the LOS, the galaxy-interloper cross-correlation changes greatly. This change can easily be understood considering that $\xi_{\rm gi}$ at $\vec{r} = \Vec{x}_1-\Vec{x}_2$, which is the separation between the two objects after misplacing interlopers by $\Delta d$, should be equal to $\xi_{\rm gi}$ at the true separation $\vec{r}\,' = \Vec{x}_1-\Vec{y}_2$. Thus,
\begin{equation}
\label{eq:cross_equation}
    \xi_{\rm gi}(\Vec{r})=\xi_{\rm g i}(\vec{r}\,') \,.
\end{equation}
There is a simple trigonometry transformation between $\vec{r}$ and $\vec{r}\,'$
\begin{equation}
\label{eq:polar_coordinates}
\begin{split}
    r'(r,\mu) = \sqrt{r^2+\Delta d^2-2 \,  \Delta d  \,  r\,  \mu} \,,
    \\
    \mu'(r,\mu) = \frac{r\mu-\Delta d}{\sqrt{r^2+\Delta d^2-2 \, \Delta d \,  r \, \mu}}\,.
\end{split}
\end{equation}
where we have defined $r= |\vec{r}|$ being the modulus of the separation vector and $\mu = {\vec{r}}\cdot\hat{z}/r$ being the cosine of the angle between the separation vector and the LOS.

The auto-correlation to cross-correlation mapping effectively pushes pair separations to larger scales. Nearby objects are more correlated than more distant objects -- the galaxy correlation function roughly decays by $r^2$, so the displacement of interlopers effectively moves the strong correlation between nearby objects around $<20\mpcoh$ to larger scales in the galaxy-interloper cross-correlation, leaving a peak. The location of the center of this peak is at the interloper displacement. We show this in Figure~\ref{fig:cross_correlation} with blue points obtained by measuring the cross-correlation between galaxies and interlopers in a Quijote simulation box at $z=1$, after misplacing interlopers by $97\mpcoh$.

In practice, if we have a small survey with higher signal-to-noise ratio (calibration data), it will be possible to estimate which galaxies would appear as interlopers in a survey with lower signal-to-noise ratio. Using calibration data we can directly measure the cross-correlation between galaxies and interlopers at their \emph{true} distance, and then infer the cross-correlation component at the wrong distance via Eq.~\ref{eq:polar_coordinates}.
If the calibration data is not available, there are two other possible approaches that we discuss in the following to estimate the cross-correlation assuming that the bias scheme of interlopers is the same as main target galaxies: analytical and observational approaches.

\emph{Analytical approach:} We considered taking the Fourier transform of the analytical power spectrum from CAMB linear and non-linear (from HALOFIT \cite{smith_stable_2003, takahashi_revising_2012}) to estimate the auto-correlation function. 
The results from the linear CAMB model are shown in red in Figure~\ref{fig:cross_correlation}. The top and bottom panels of this plot show that the linear CAMB model roughly describes the shape of the measured monople and quadrupole of the cross-correlation, but it underestimates the peak height by $\sim 30\%$. This is because the prediction of CAMB for the monopole at small scales is significantly lower than the non-linear reality, and, as we discussed earlier, this directly impacts the height of the peak at $97 \mpcoh$. We have also used HALOFIT to create a non-linear model. The corresponding results are shown in green in the same figure. The outcome remains the same with a slightly better prediction of the height. In this case, while the displacement depends on the fiducial cosmology and it determines the position of the peak in the target-interloper correlation function, the shape of the cross-correlation depends on the true cosmology since it is a remapping of the auto-correlation of target galaxies via Eq.~\ref{eq:Cartesian_to_polar}. This can be easily implemented in the model for the correlation function. The inability of this model to accurately fit the cross-correlation function led us to consider our second approach, or the observational approach.

\emph{Observational approach:} 
If we assume that interlopers and targets are drawn from the same population of galaxies, we can estimate their cross-correlation directly from the measured auto-correlation of the contaminated catalogue As shown in Figure~\ref{fig:cross_correlation}, the monopole of cross-term vanishes on small scales ${r}<40\mpcoh$. Therefore, one can rewrite Eq.~\ref{eq:xi_tt} as
\begin{equation}
    \label{eq:xi_tt_approx}
    \xi(\Vec{r};f_{\rm i}) \approx (1+2f_{\rm i}^2-2f_{\rm i})\,\xi_{\rm gg}(\Vec{r}) \quad {\rm for \, small \, r,}
\end{equation}
indicating that the contaminated and uncontaminated catalogs have small-scale correlation functions that differ by an overall amplitude only, and this amplitude depends on the fraction of interlopers in the contaminated catalogue We can therefore infer the galaxy-galaxy small-scale correlation function directly from the measured contaminated correlation function, modulo an overall normalization. The large-scale feature in the cross-correlation seen in Figure~\ref{fig:cross_correlation} is mainly determined by the small-scale correlation of galaxies and interlopers at the true position, or equivalently the galaxy auto-correlation since we assumed that galaxies and interlopers have the same bias (see Eq.~\ref{eq:cross_equation}). Therefore, using Eqs.~\ref{eq:xi_tt_approx} and~\ref{eq:cross_equation} we can estimate the cross-correlation term on the scales of interest ($r<150\mpcoh$) directly from the measured small-scale contaminated correlation function as 
\begin{equation}
    \label{eq:cross_equation_approx}
    \xi_{\rm gi}(\Vec{r})=\frac{\xi(\vec{r}\,';f_{\rm i})}{{(1+2f_{\rm i}^2-2f_{\rm i})}} \,,
\end{equation}
where the mapping between $\Vec{r}$ and $\vec{r}\,'$ is described in Eq.~\ref{eq:polar_coordinates}. Using this observational method, the model for the cross-correlation is already built from a correlation function in the true cosmology but measured in the fiducial one, and we will not need to account for differences between the two cosmologies when performing the fit.

Using this method to determine the cross-correlation, one can immediately see the improvement in the peak height estimation of the monopole and quadrupole moments of the cross-correlation function in Figure~\ref{fig:cross_correlation} (orange line). Note that this method assumes that interlopers and targets have the same bias scheme, and this assumption is satisfied exactly in our mocks. 
On the other hand, the general shape of the monopole of the cross-correlation, especially at larger scales, deviates from that of the measurement. We believe that this is due to the approximation that we made to derive Eq.~\ref{eq:xi_tt_approx}; that of a vanishing cross-correlation on small scales. The quadrupole is less affected by this approximation. These effects do not alter the measured parameters (see Section~\ref{subsec:Model_and_Parameters}). Finally, using this approximation, we can use Eq.\ref{eq:cross_equation_approx} for the galaxy-interloper cross-correlation and rewrite Eq.~\ref{eq:xi_tt}
 
\begin{equation}
    \label{eq:xi_final}
    \xi(\Vec{r};f_{\rm i}) = (1+2f_{\rm i}^2-2f_{\rm i})\xi_{\rm gg}(\Vec{r})+\frac{2f_{\rm i}(1-f_{\rm i})}{(1+2f_{\rm i}^2-2f_{\rm i})}
   \times \mathcal{M}[\xi(f_i)](\vec{r})\,,
\end{equation}
where for simplicity, we use $\mathcal{M}$ to denote the mapping of scales in Eq.~\ref{eq:polar_coordinates} that describes the deformation due to interlopers. 

\subsection{Building a Cosmological Model}\label{subsec:Model_and_Parameters}
Thus far, we have introduced a model for the contaminated correlation function that consists of the galaxy-galaxy correlation function term, $\xi_{\rm gg}(\Vec{r})$, and the cross-correlation term, $\xi_{\rm gi}(\Vec{r};f_{\rm i})$. 
There is still one part missing: How does our model depend on the assumptions about the fiducial cosmology? In this Section, we will address this question based on the methodology and notation of Ref.~\cite{Padmanabhan_White_2008} and \cite{Xu_2013}.

To measure the correlation function from data, we need to calculate the separation between pairs by assuming a fiducial cosmology to translate from redshifts to distances. The BAO peak in the correlation function can be distorted if the assumed cosmology is not the same as the true cosmology. This distortion can be
parameterized by Alcock–Paczynski parameters (hereafter AP; \cite{alcock_evolution_1979}): $\alpha$, the isotropic dilation parameter, and $\epsilon$, the anisotropic warping parameter:
\begin{equation}
    \begin{split}
        \alpha = \left[
        \frac{D_{\rm A, \rm true}^2(z) H_{\rm fid}(z)}{D_{\rm A,  fid}^2(z) H_{\rm true}(z)}
        \right]^{1/3}
        \frac{r_{\rm s,fid}}{r_{\rm s, true}}\,,\\
        1 + \epsilon = \left[\frac{D_{\rm A, fid}(z) H_{\rm fid}(z)}{D_{\rm A, true}(z) H_{\rm true}(z)}\right]^{1/3}\,.
    \end{split}
\end{equation}
And we can relate the fiducial and true distances using $\alpha$ and $\epsilon$ using the following equations:
\begin{equation}
    \begin{split}
        r_{\parallel, \rm true} = \alpha (1+\epsilon)^2 r_{\parallel}\,, \\
        r_{\perp, \rm true} = \alpha (1+\epsilon)^{-1} r_{\perp}\,,
    \end{split}
\end{equation}
where we have dropped subscript "fid" for simplicity. The separation between two objects can be described in Cartesian $(r_{\perp}, r_{\parallel})$ or polar coordinates $(r, \mu)$, which are related by
\begin{equation}
    \label{eq:Cartesian_to_polar}
    \begin{split}
        r^2 = r_{\perp}^2 + r_{\parallel}^2\,,\\
        \mu = {\rm cos(\theta)} = \frac{r_{\parallel}}{r}\,.
    \end{split}
\end{equation}
Therefore Eq.~\ref{eq:Cartesian_to_polar} for the true separations would be
\begin{equation}
    \begin{split}
        r_{\rm true} = \alpha \sqrt{(1+\epsilon)^4 r_{\perp}^2 +(1+\epsilon)^{-2} r_{\parallel}^2}\\
        \mu_{\rm true} = \frac{ (1+\epsilon)^2 r_{\parallel}}{\sqrt{(1+\epsilon)^4 r_{\perp}^2 +(1+\epsilon)^{-2} r_{\parallel}^2}}
    \end{split}
\end{equation}

The true correlation function can be decomposed into its Legendre multipoles. In the following, we assume that monopole, quadrupole, and hexadecapole would suffice in the summation below, as the higher order moments can be ignored.
\begin{equation}\label{eq:Legendre}
    \xi(r_{\rm true}, \mu_{\rm true})
    = 
    \sum_{l = 0,2,4} \xi_{l}(r_{\rm true}) \,  \mathcal{L}_{l}(\mu_{\rm true})
\end{equation}
What we measure in reality, assuming a wrong fiducial cosmology, is shifted multipole moments of the correlation function\,,
\begin{equation}
    \xi_{l} (r; \alpha,\epsilon)= \frac{2l+1}{2}\int_{-1}^{1} d\mu L_l(\mu)\, \xi(r_{\rm true}(\alpha,\epsilon),\mu_{\rm true}(\alpha,\epsilon))\,.
\end{equation}

For a BAO-only measurement, we are not interested in any feature of the correlation function other than where the BAO is. Typically, one adds an additive polynomial term $A(r)$ to account for the scale-dependent bias and redshift space distortions, and a multiplicative $B(r)$ factor to adjust the amplitude,
\begin{equation}\label{eq:xi_model_uncontaminate}
    \xi_{l}^{\rm model}(r; \alpha, \epsilon) = B(r)\xi_{l}(r; \alpha, \epsilon) + A(r))
\end{equation}
Using the right form of polynomials is important, because even though $A(r)$ and $B(r)$ are nuisance parameters not containing any BAO information, they may bias our estimation of the BAO peak if not properly fitted for. We use three orders of polynomials: $A(r) = a_1/r^2 + a_2/r + a_3$, and a constant for the amplitude $B(r) = B_0$.
Substituting Eq.~\ref{eq:xi_model_uncontaminate} into Eq.~\ref{eq:xi_final} gives our final model for the monopole and quadrupole of a catalogue contaminated by some fraction of interlopers $f_{\rm i}$

\begin{equation}
\begin{split}
    \label{eq:xi_model_0}
    \xi_{0}^{\rm model}(r;\{a_1, a_2, a_3, B_0, \alpha, \epsilon, f_{\rm i}\}) \\
    &=
    (1+2f_{\rm i}^2-2f_{\rm i})
    \times 
    \left[
    (\frac{a_1}{r^2} + \frac{a_2}{r} + a_3) +
    B_0 \, \xi^{\rm}_{0}(r; \alpha,\epsilon))
    \right]
    \\
    &+
    \frac{2f_{\rm i}(1-f_{\rm i})}{(1+2f_{\rm i}^2-2f_{\rm i})}
    \times \mathcal{M}_{0}[\xi^{\rm measured}](r)\,,
\end{split}
\end{equation}

\begin{equation}
\begin{split}
    \label{eq:xi_model_2}
    \xi_{2}^{\rm model}(r;\{w_1, w_2, w_3, B_0, \alpha, \epsilon, f_{\rm i}\}) \\
    &=
    (1+2f_{\rm i}^2-2f_{\rm i})
    \times 
    \left[
    (\frac{w_1}{r^2} + \frac{w_2}{r} + w_3) +
    B_0 \, \xi^{\rm}_{2}(r; \alpha,\epsilon))
    \right]
    \\
    &+
    \frac{2f_{\rm i}(1-f_{\rm i})}{(1+2f_{\rm i}^2-2f_{\rm i})}
    \times \mathcal{M}_{2}[\xi^{\rm measured}](r)\,.
\end{split}
\end{equation}

Throughout this paper, when we refer to a monopole only fit, we use Eq.~\ref{eq:xi_model_0} assuming $\epsilon = 0$, which simplifies $\xi^{\rm}_{0}(r; \alpha,\epsilon)$ to just $\xi^{\rm}_{0}(\alpha r)$. In that case, there are six free parameters that we fit for. When we refer to simultaneous monopole and quadrupole fit, we use Eq.~\ref{eq:xi_model_0} and \ref{eq:xi_model_2} together. In that case, there are four additional free parameters (ten in total) that should be fitted for.

In the following, we describe the method used to obtain the templates of the monopole and quadrupole, $\xi_0(r)$ and $\xi_2(r)$. 
We model the matter power spectrum in real space as
\begin{equation}
    P(k,\mu) = \left[ P_{\rm lin}(k) - P_{\rm smooth}(k)\right] e^{-\frac{1}{2} k^2 (\mu^2 \Sigma_{\parallel}^2 + (1- \mu^2) \Sigma_{\perp}^2)} + P_{\rm smooth}(k)\,.
\end{equation}
where $P_{\rm lin}(k)$ is the linear matter power spectrum obtained from CAMB and $P_{\rm smooth}(k)$ is the
no-wiggle counterpart computed as in \cite{vlah_perturbation_2016}, where the BAO oscillations have been removed. The Gaussian term accounts for the damping of the BAO wiggles due to non-linear evolution, with $\Sigma_{\perp}=4.8\mpcoh$ and $\Sigma_{\parallel}=7.3 \mpcoh$ at $z=1$ and the considered cosmology. In redshift space, the power spectrum is distorted by the Kaiser \cite{kaiser_clustering_1987} and the Finger-of-God (FoG) \cite{jackson_critique_1972} effects, which lead us to the following equation for the redshift space power spectrum \cite{blake_wigglez_2011}
\begin{equation}
    P_{\rm s}(k,\mu) = b^2 (1+\beta \mu^2)^2 F(k,\mu)P(k,\mu)\,.
\end{equation}
In our analysis, we use the streaming model for FoG, $F(k,\mu) = (1+k^2\mu^2\Sigma_s^2)^{-2}$,  with $\Sigma_{s} = 3 \mpcoh$.
Subsequently, we calculate the correlation function by taking the Fourier transform of the redshift-space power spectrum.
\begin{figure}[t]
    \centering
    \includegraphics[width=1\textwidth,angle=0]{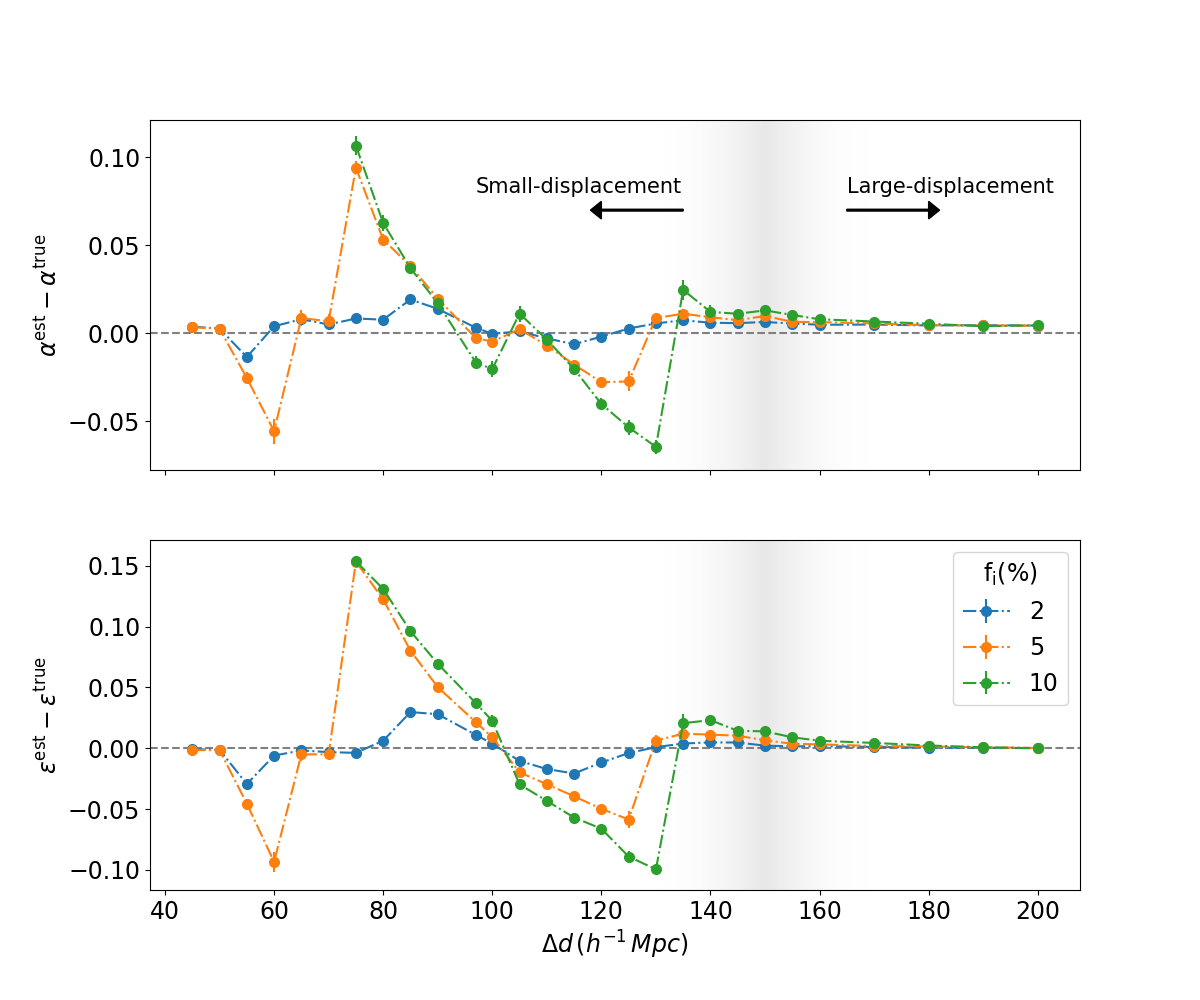}
    \caption{
    The effect of small-displacement interlopers on dilation parameters when interlopers are not included in the model, shown for $2$, $5$, and $10\%$ fraction of interlopers. As can be seen, for small-displacements, the estimation of $\alpha$ (upper panel) and $\epsilon$ (lower panel) are notably affected by interlopers, and therefore, small-displacement interloper corrections need to be implemented in the fitting model to avoid systematic bias.
    For large-displacements, the effect of the cross-correlation on the recovered parameters rapidly decreases.
    }
    \label{fig:dilation_vs_deltad}
\end{figure}

\section{Results}
\label{sec:results}
\begin{figure}[t]
    \centering
    \includegraphics[width=1\textwidth,angle=0]{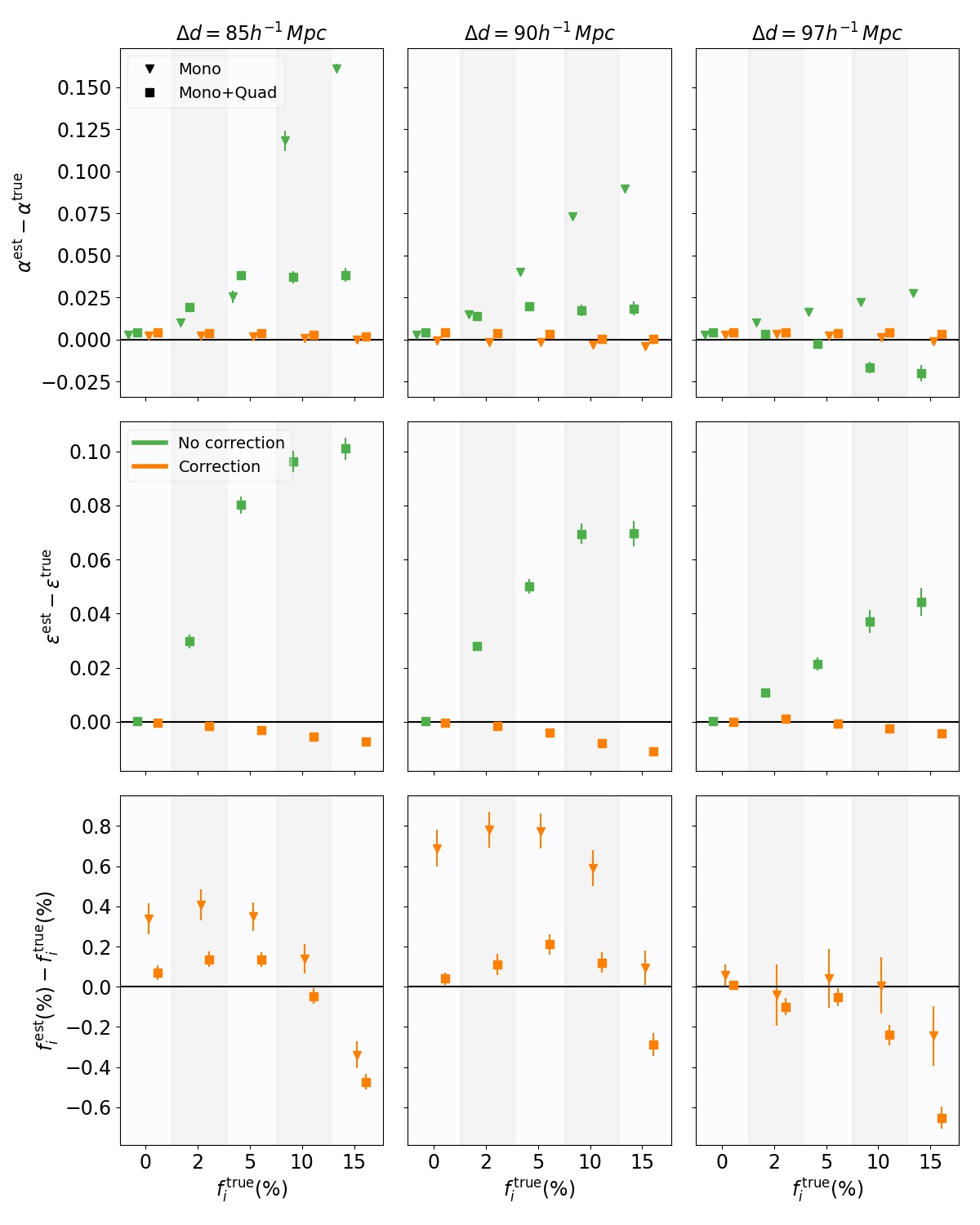}
    \caption{
    The difference between estimation and true for $\alpha$ (top row), $\epsilon$ (middle row), and $f_i$ (bottom row), as plotted against the true fraction of interlopers. For comparison, the green lines show the fit without taking into account interlopers and the orange shows the fit with interlopers (our analysis). 
    } 
    \label{fig:results}
\end{figure}

In this Section, we present the results obtained by fitting the mean of the multipole moments of the correlation functions from $1,000$ mocks in redshift space.
To find the best fit parameters, we carried out a Markov-chain Monte Carlo (hereafter, MCMC) analysis (emcee package \cite{foreman_mackey_emcee_2013}) to minimize the posterior function introduced by Percival et al. \cite{Percival_posterior2021}:
\begin{equation}
\label{eq:posterior}
    f(\xi^{\rm model} |\xi^{\rm data},C) \propto \left[ 1+\frac{\chi^2}{n_s-1} \right]^{-\frac{m}{2}}\,,
\end{equation}
where $m$ is given by Eq.~54 of \cite{Percival_posterior2021}. 
We also assumed the following uniform priors on our parameters:
\begin{equation}
  f(\alpha, \epsilon, f_i) =
    \begin{cases}
      1 & \text{if $0.6<\alpha<1.4$}\,,\\
      & \text{\& $-1<\epsilon<1$}\,,\\
      & \text{\& $0<f_i<0.5$}\\
      0 & \text{otherwise}
    \end{cases}       
\end{equation}
The best fit values and the statistical errors presented in this Section (i.e., Figure~\ref{fig:dilation_vs_deltad}, \ref{fig:results}, Table~\ref{tab:results_all}, \ref{tab:results_mono_only}) are respectively calculated from the mean and standard deviation calculated using a Monte Carlo Markov Chain used to explore the posterior with 3000 steps.
To calculate $\chi^2$, we estimated the covariance matrix of $n_s = 1000$ independent contaminated mocks using
\begin{equation}
\label{eq:covariance}
    C_{ij}[\xi(r_i)\xi(r_j)] = \frac{1}{(n_s-1)}\sum_{n=1}^{n_s} 
    [\xi_n(r_i) - \bar{\xi}(r_i)]
    [\xi_n(r_j) - \bar{\xi}(r_j)]\,.
\end{equation}
To fit the model to the mean of $n_s$ mocks, Eq.~\ref{eq:covariance}, which gives the covariance matrix used for an individual mock, is divided by $n_s$. For the purpose of this paper, we only need to fit the data around the BAO peak. Therefore, in our analyses, we fit our model to data on scales between $r_{\rm min} = 50 \mpcoh$ and $r_{\rm max} = 150 \mpcoh$. 

Let us begin with highlighting the necessity of correcting for the small-displacement interloper effect, and also roughly quantifying what we mean by "small" displacements. For this purpose, we repeated the process of making contaminated catalogues described in Section~\ref{sec:simulations}, but for a wider range of displacements, from $45$ to $200 \mpcoh$. Without correcting for the interloper effect, we found the best fit estimations on $\alpha$ and $\epsilon$. In Figure~\ref{fig:dilation_vs_deltad}, the difference between the estimation and true is plotted against displacement, and from this figure we find that:

\begin{itemize}
    \item How much the estimated dilation parameters deviate from their true value depends on the level of contamination of the catalogue (different colors). In the small-displacement regime, estimations are worse for higher fractions of of interlopers. For $10\%$ interlopers, the MCMC code does not even converge for displacements smaller than $75 \mpcoh$ (hence why not shown in this plot).
    \item 
    For $\Delta d \approx 100 \mpcoh$, the best fit does not differ from 1 significantly, since the two peaks ($\Delta d$ and BAO) overlap, and there is no need to change the peak location in the model by varying $\alpha$. As we make $\Delta d$ slightly larger (smaller) than the BAO scale, until around $120 \mpcoh$ ($80 \mpcoh$), $\alpha$ becomes smaller (larger) than 1 to allow the model without the interloper signal to compromise between fitting the BAO and the interloper cross-correlation term. 
    \item 
    There is a smooth transition around $\Delta d \approx 140-160 \mpcoh$ such that, at smaller displacements, the estimated dilation parameters deviate from the truth, while on larger scales they become less affected by the presence of interlopers. 
    This is because as we increase the displacement, the cross-correlation peak phases out of the maximum fitting range of $150 \mpcoh$, and the tail that remains within the fitting range gets corrected by polynomials, rather than the dilation parameters. 
    \item 
    The extrema reflect a jump in the best-fit position in the posterior between two isolated solutions with different $\Delta d$: when the cross-correlation becomes too far from the position of the BAO, the best fit solution jumps from being a compromise between fitting both BAO and spike to a solution that fits the BAO (only) better, ignoring the cross-correlation spike.
\end{itemize}

We should emphasize that large-displacements can still bias the BAO measurement, even though the cross-correlation would be negligible, as mentioned in Section~\ref{sec:intro}.

Now, we take into account the effect of interlopers in the fitting model by using the formalism described in Section~\ref{sec:Modelling_the_contaminated_correlation_function}. Figure~\ref{fig:results} demonstrates how much the estimation of our pipeline for $\alpha$ (top row), $\epsilon$ (middle row), and $f_i$ (lower row) deviates from the corresponding true values when considering catalogs contaminated by different fractions of interlopers (x-axis), and three different values for the displacements $\Delta d$: $85\mpcoh$ (left column), $90\mpcoh$ (middle column), and  $97 \mpcoh$ (right column). The orange symbols corresponds to the case where we account for the presence of interlopers in our analysis using Eq.~\ref{eq:xi_final}.
We also consider green symbols which show the results when ignoring interlopers (as if $f_i$ was set to zero in Eq.~\ref{eq:xi_final}, and as assumed in Figure~\ref{fig:dilation_vs_deltad}). We considered fits to the monopole alone (shown in triangle) and to the monopole and quadrupole simultaneous (rectangles). From Figure~\ref{fig:results}, we deduce that:
\begin{itemize}
    \item 
    \emph{Monopole only with no correction (green triangles):}
    At a fixed value for the displacement $\Delta d$, the estimated isotropic dilation parameter $\alpha_{\rm est}$ increasingly deviates from 1 when considering larger interloper fractions. This means that the bias in $\alpha$ introduced by the interlopers increases with the amount of contamination, as previously noted in \cite{massara2020line}. How much the bias increases with $f_i$ depends on the value of the displacement: the smaller the displacement, the faster the bias increases with the amount of contamination. Indeed, when the displacement is close to the BAO position, e.g. $\Delta d = 97\mpcoh$, the peak in the galaxy-interloper cross-correlation is very close to the BAO location and it causes a mild distortion and shift of the BAO peak. On the other hand, smaller displacements present a peak in the cross-correlation that is further away from the BAO peak location, causing a much larger distortion of the BAO feature (see also Figure~\ref{fig:mono_general}).  
    \item 
    \emph{Monopole+Quadrupole with no correction (green rectangles):}
    Compared to the previous case (monopole only with no correction) and for each combination of $\Delta d$ and $f_i^{\rm true}$, $\alpha_{\rm est}$ is closer to $1$ when including the quadrupole in the fit. This happens because the contamination is anisotropic: By including the quadrupole, the parameter $\epsilon$ can mimic and absorb the anisotropy, leaving $\alpha$ less biased. However, the $\epsilon$ measurements are strongly biased, with deviations from the fiducial value increasing with the amount of contamination and decreasing with increasing $\Delta d$.

    \item 
    \emph{Monopole only with interloper correction (orange triangles):}
    Compared to the no correction case, the values of $\alpha_{\rm est}$ are now much closer to the true value and they exhibit a systematic bias smaller than $3.9 \times 10^{-3}$ for all considered level of contamination and displacements. This model allows us to obtain an estimation for the fraction of interlopers $f_i^{\rm est}$ with a bias that is below $8 \times 10^{-3}$ for all levels of contamination and displacements. (More details in Table~\ref{tab:results_mono_only})
    \item 
    \emph{Monopole+Quadrupole with interloper correction (orange rectangles):} The values of $\alpha_{\rm est}$ and $\epsilon_{\rm est}$ are much closer to the truth than in the no correction case. The residual systematic biases are smaller than $5 \times 10^{-3}$ for $\alpha$ and $0.01$ for $\epsilon$. The percentage of contaminants can be estimated with better precision in this case than when fitting for the monopole alone: the statistical errors decrease by a factor of two on average. The accuracy on $f_i$ is also generally improved when the displacement is small, while no significant amelioration appears if $\Delta d = 97\mpcoh$.
\end{itemize}

A more detailed description of our results of the monopole+quadrupole and monopole only analyses are given in Table~\ref{tab:results_all} and Table~\ref{tab:results_mono_only} respectively, where the systematic and statistical errors on $\alpha$, $\epsilon$ and $f_i$ are reported for all the displacements and fractions considered. We find that the systematic error on $\alpha$ with interlopers using our model is $4.4 \times 10^{-3}$ which is consistent with pre-reconstruction BAO analyses on SDSS data without interlopers (\cite{bautista_completed_2020}, \cite{magana_sdss-iii_2013}, \cite{alam_clustering_2017}). This shows that slitless spectroscopy does not degrade BAO measurements. 

\begin{center}

\begin{table}[t]
\caption{\label{tab:results_all} 
Systematic and statistical errors on $\alpha$, $\epsilon$, and $f_i$ using monopole and quadrupole. Analysis is performed on the mean of 1000 mocks in redshift space, and different displacements.\\}
\centering
 \begin{tabular}{|| c | c c | c c | c c ||} 
\hline\hline
$f_i^{\rm true} (\times 10^{-2})$&
\multicolumn{2}{c|}{Error on $\alpha$ $(\times 10^{-3})$}&
\multicolumn{2}{c|}{Error on $\epsilon$ $(\times 10^{-3})$}&
\multicolumn{2}{c||}{Error on $f_i$ $(\times 10^{-3})$}\\

 &
 systematic &
 statistical &
 systematic &
 statistical &
 systematic &
 statistical 
 \\
 
\hline\hline
\multicolumn{7}{c}{$\Delta d = 85 \rm h^{-1}Mpc$}\\
\hline\hline

0 & 4.1 & 0.7 & -0.5 & 0.9 & 0.7 & 0.4\\
2 & 3.9 & 0.8 & -1.5 & 1.1 & 1.4 & 0.4\\
5 & 3.5 & 0.8 & -3.1 & 1.1 & 1.4 & 0.4\\
10 & 2.5 & 0.8 & -5.4 & 1.0 & -0.5 & 0.4\\
15 & 1.6 & 0.9 & -7.3 & 1.1 & -4.7 & 0.4\\

\hline\hline
\multicolumn{7}{c}{$\Delta d = 90 \rm h^{-1}Mpc$}\\
\hline\hline

0 & 4.1 & 0.7 & -0.5 & 1.0 & 0.4 & 0.3\\
2 & 3.6 & 0.9 & -1.5 & 1.4 & 1.1 & 0.5\\
5 & 3.0 & 0.8 & -4.1 & 1.3 & 2.1 & 0.5\\
10 & 0.3 & 0.8 & -7.9 & 1.3 & 1.2 & 0.5\\
15 & 0.3 & 0.9 & -10.8 & 1.5 & -2.9 & 0.6\\

\hline\hline
\multicolumn{7}{c}{$\Delta d = 97 \rm h^{-1}Mpc$}\\
\hline\hline
 
0 & 4.3 & 0.7 & 0.1 & 0.8 & 0.1 & 0.1\\
2 & 4.4 & 0.7 & 1.1 & 0.9 & -1.0 & 0.4\\
5 & 3.5 & 0.8 & -0.7 & 1.1 & -0.5 & 0.5\\
10 & 3.9 & 1.1 & -2.5 & 1.5 & -2.4 & 0.5\\
15 & 3.4 & 1.2 & -4.3 & 1.7 & -6.5 & 0.6\\

\hline\hline
\end{tabular}
\end{table}
\end{center}

\begin{center}

\begin{table}[t]
\caption{\label{tab:results_mono_only} 
Systematic and statistical errors on $\alpha$, and $f_i$ using the monopole only analysis performed on the mean of 1000 mocks in redshift space, and different displacements.\\}
\centering
 \begin{tabular}{|| c | c c | c c ||} 
\hline\hline
$f_i^{\rm true} (\times 10^{-2})$&
\multicolumn{2}{c|}{Error on $\alpha$ $(\times 10^{-3})$}&
\multicolumn{2}{c||}{Error on $f_i$ $(\times 10^{-3})$}\\

 &
 systematic &
 statistical &
 systematic &
 statistical 
 \\

\hline\hline
\multicolumn{5}{c}{$\Delta d = 85 \rm h^{-1}Mpc$}\\
\hline\hline

0 & 2.2 & 0.7 &  3.4 & 0.8\\
2 & 2.1 & 0.7 & 4.1 & 0.8\\
5 & 1.9 & 0.7 & 3.5 & 0.7\\
10 & 0.8 & 0.7 &  1.4 & 0.7\\
15 & -0.1 & 0.8 & -3.4 & 0.7\\

\hline\hline
\multicolumn{5}{c}{$\Delta d = 90 \rm h^{-1}Mpc$}\\
\hline\hline

0 & -0.6 & 0.8 & 6.9 & 0.9\\
2 & -1.4 & 0.8 &  7.8 & 0.9\\
5 & -1.5 & 0.9 &  7.7 & 0.9\\
10 & -3.0 & 0.9 & 5.9 & 0.9\\
15 & -3.9 & 0.9 & 1.0 & 0.8\\

\hline\hline
\multicolumn{5}{c}{$\Delta d = 97 \rm h^{-1}Mpc$}\\
\hline\hline
 
0 & 2.7 & 0.7 & 0.6 & 0.5\\
2 & 3.2 & 0.9 & -0.4 & 1.5\\
5 & 2.4 & 0.9 &  0.4 & 1.5\\
10 & 1.3 & 1.0 &  0.1 & 1.4\\
15 & -0.9 & 1.2 & -2.4 & 1.5\\

\hline\hline
\end{tabular}
\end{table}
\end{center}

\section{Discussions}\label{sec:discussions}

We have investigated how to extract the BAO scale from a contaminated galaxy catalogue containing a sub-sample of interlopers that are correlated with the main target galaxies. This case is particularly important for the future Roman Space Telescope, where a population of \oiii\ emitting galaxies extracted thanks to the observation of a single line is likely to be contaminated by $\rm H\beta$ interlopers. 
In this study, we assumed that the $\rm H \beta$ line is misidentified as the primary line at $500.7 \, \rm nm$ of the \oiii \ doublet.
Depending on how the doublet is fitted, one might need to take into account the weighted average of the primary and the secondary lines of \oiii. Nonetheless, this will only impact the value of $\Delta d$ as a function of redshift in Eq.~\ref{eq:displacement}, and will not affect our conclusions.
In the following, we highlight some of the main outcomes of our analysis.

Generally, the contaminated correlation function has three terms (Eq.\ref{eq:xi_tt_general}): The galaxy-galaxy term, the interloper-interloper term, and the galaxy-interloper term. We have split interlopers into three categories: random, large-displacement and small-displacement interlopers. We have argued that random contaminants, uncorrelated with either themselves or the target galaxies are trivial to account for.

For both small-displacement and large-displacement interlopers, we need to model their auto-correlation, but this is generally easy to do. In addition, the effect of the interloper-interloper auto-correlation on the contaminated signal is a simple linear super-position of two correlation functions: in the extremes, either the change in $H(z)$ between true redshifts of the targets and interlopers is small and so the BAO position is unaffected, or it is large and the superimposed BAO peak is far from that of the targets. In either case and in between, to measure the BAO position, we only need a model on large scales where linear theory is sufficient to describe the BAO feature and we know the position through the relative line positions. If interlopers and galaxies have different bias, an extra (bias) parameter can be added to the second term of Eq.~\ref{eq:xi_tt_general} and can be fitted for along other parameters introduced in our analysis. It also may be possible to measure this using calibration data, by comparing the interloper-interloper and target-target correlation functions. In contrast, for small-displacement interlopers, it is critical to have a good estimate of the interloper-target cross-correlation. 

For our simulations, the interlopers are drawn from the same underlying distribution as the target galaxies. This allowed us to simplify the equations, merging the galaxy-galaxy and interloper-interloper terms together. In practice, the populations of \oiii\ and $\rm H \beta$ emission line galaxies are different, and this can change our equations requiring us to keep these terms separate. We do not expect this to significantly alter our conclusions, or to present a difficulty for modeling. 

For small-displacement interlopers, the interloper shift can be considered as being constant, and is dependent on the fiducial cosmology used to convert redshifts to distances. Thus its position does not provide cosmological information. We made the assumption of a constant shift because the displacement only changes by less than one percent across the bin considered, similar to the case for the BAO scale being fixed within an bin of observational data. Hence, the BAO signal in $\xi_{\rm ii}(\Vec{r})$ will not be shifted as a correlation function is not modified by a translation, while it is modified in $\xi_{\rm gi}(\Vec{r})$.
Thus, the key to understanding and modelling the contaminated correlation function is through accurately modelling the cross-correlation between the main galaxy targets and the interlopers that are misidentified as main targets, which is described in Eq.~\ref{eq:cross_equation} and required a mapping between true and wrongly inferred positions for the interlopers.
Again, with calibration data, it is possible to apply this mapping (refer to \ref{eq:polar_coordinates}) to the measured cross-correlation at the true distance. 
We leave it to future work to determine the size of calibration data required to determine this function with sufficient accuracy that it does not impact on BAO measurements. The good news is that we need to know this function on small scales, where small volumes of the Universe will be sufficient. 

For our study, we used the same assumptions used to create the simulations (the \oiii\ and $\rm H \beta$ emitting galaxies are drawn from the same population) to estimate the cross-correlation with the auto-correlation function of the main galaxy targets. We considered models based on CAMB and HALOFIT to derive an estimation of the cross-correlation. We found that they both fail in precisely modelling the peak of the monopole and quadrupole of the cross-correlation. These models also need to account for the difference between fiducial and true cosmologies via the AP parameters, that would need to be added in the cross-correlation function of Eq.~\ref{eq:xi_model_0}, and Eq.~\ref{eq:xi_model_2} as well.
Therefore, we tested another possibility that is more powerful on smaller scales: using the measured contaminated correlation functions to estimate cross-correlation. We argued that using the latter is the most precise for the Quijote mocks used in this paper, since it matches the way that the mocks were constructed (interlopers and target galaxies have the same galaxy bias). Indeed, the resulting cross-correlation is able to better model the peak of the monopole of the cross-correlation, and both peak and the shape of the quadrupole of the cross-correlation (refer to Figure~\ref{fig:cross_correlation}).

With all of the mentioned assumptions made, we finally presented a model for monopole and quadrupole of the contaminated correlation function. We tested our model by fitting it to the mean of 1,000 mocks for five different interloper fractions and three different displacements. We found that for all these cases, our method is successful at making an unbiased prediction of the dilation parameters (recovering the BAO peak), with a systematic error of less than $4.4 \times 10^{-3}$, which is consistent with previous pre-reconstruction analyses. It also enables us to have a good estimation of the fraction of interlopers. Whereas if we did not take interlopers into account in the model, for $15\%$ interlopers we would have an order of magnitude higher systematic errors -- as high as $-0.020$ and $0.044$ for $\alpha$ and $\epsilon$. 
Moreover, the signal in the correlation function from interlopers is sufficiently clear that it allows us to measure the fraction of interlopers with a systematic error of at most few percent. This will be useful for cosmological analysis other than measuring BAO where the fraction of interlopers can change the amplitude of the signal, and knowing the fraction is important.

Since our analysis is specifically useful for future Roman \oiii\ targets that are contaminated by $\rm H \beta$ interlopers, we predict the statistical error that our method would have for Roman telescope data. To do so, we scaled the statistical errors on the mean calculations from 1,000 mocks at redshift 1 ($\Delta d = 85 \mpcoh$). These 1,000 mocks have a cumulative volume of $\rm 1,000 (h^{-1} Gpc)^3$. The approximate volume of Roman survey between redshift 1.8 and 2.8 (where \oiii\ galaxies live) is $V \approx \rm 10 (h^{-1} Gpc)^3$. Thus, the statistical error is scaled by $\sqrt{1000/10} = 100$.
For the range of $0-15\%$ interlopers, we find that the statistical error on $\alpha$ and $\epsilon$ is $0.01$ which is larger than their corresponding systematic errors.

Small-displacement interlopers give rise to a strong signal in the correlation function on similar scales to the cosmological signal. Even so, this signal has key differences from the cosmological signal such that a joint model fitting both the cosmological signal and the effect of interlopers can be constructed and we do not see a degradation in the cosmological measurements in the tests we have performed. Furthermore, the interloper signal itself allows excellent diagnostics on the contamination, allowing a simultaneous fit to the contaminant fraction. For random interlopers however, it is more difficult to measure the contaminant fraction, which has to be done before the amplitude of clustering can be measured. Thus, while small-displacement interlopers are definitely the most pernicious in terms of the correlation function, for fits to the data this is not the case. 
\acknowledgments

We thank Yun Wang for useful comments and discussions. Research at Perimeter Institute is supported in part by the Government of Canada through the Department of Innovation, Science and Economic Development Canada and by the Province of Ontario through the Ministry of Colleges and Universities. This research was enabled in part by support provided by Compute Ontario (computeontario.ca) and the Digital Research Alliance of Canada (alliancecan.ca). 


\printbibliography
\end{document}